\documentclass[a4paper,12pt]{article}

\usepackage{amsmath,amssymb,amsfonts}

\usepackage{url}

\usepackage{float} 

\usepackage{graphicx}
\usepackage[usenames,dvipsnames]{xcolor}
\usepackage{epsfig}
\usepackage{epstopdf}
\usepackage{dcolumn}
\usepackage{tikz}
\usetikzlibrary{shapes.geometric, arrows}
\usepackage{upgreek}
\usepackage{setspace}
\usepackage{enumitem}
\usepackage{array,multirow,bigdelim,arydshln}
\usepackage{bbold}
\usepackage{color}

\numberwithin{equation}{section}

    \def\be{\begin{equation}}
    \def\ee{\end{equation}}
    \def\pb{\bar\partial}

    \def\({\left(}
    \def\){\right)}
    \def\<{\left\langle\,}
    \def\>{\, \right\rangle}
    \def\[{\left[}
    \def\]{\right]}

    \def\Im{{\rm Im ~}}
\def\a{\alpha}
\def\b{\beta}
\def\g{\gamma}
\def\l{\lambda}
\def\o{\omega}

\def\lb{\bar\lambda}
\def\de{\delta}
\def\r{\rho}
\def\e{\epsilon}
\def\t{\theta}
\def\k{\kappa}
    \def\p{\partial}

   \def\-{^}

\def\g{\gamma}

\def\r{\rho}

\def\s{\sigma}

\def\l{\lambda}

\def\o{\omega}

\def\t{\theta}

\def\G{\Gamma}

\def\L{\Lambda}

\def\p{\partial}
\def\pb{\overline\partial}

\def\pp{\prime}

\begin{document}

    \vskip1.1truein
    \centerline{ }
    \vskip 2.2cm
    
    \centerline{\Large{\bf An Introduction to Pure Spinor Superstring Theory}} 
   % \vskip 0.27cm
   %\centerline{\Large{\bf Integrals on Pure
    % Spinors Space}}
    \vskip 1.3cm
     \centerline{Nathan Berkovits$^{\dagger,}$\footnote{nberkovi@ift.unesp.br } and Humberto
Gomez$^{*,}$\footnote{humgomzu@gmail.com}}
\vskip 25pt
 \centerline{\sl 
$^{\dagger}$
ICTP South American Institute for Fundamental Research}
 \centerline{\sl 
Instituto de F\'isica Te\'orica
UNESP - Universidade Estadual Paulista}
\centerline{\sl Rua Dr. Bento T. Ferraz 271, 01140-070, S\~ao Paulo, SP, Brasil.}    
    \vskip 25pt  
    \centerline{\sl 
$^*$Facultad de Ciencias Basicas,  Universidad Santiago de Cali,}
    \centerline{\sl 
Calle 5 $N^\circ$  62-00 Barrio Pampalinda, Cali, Valle, Colombia.}

     \vskip 1.3cm
    
\begin{abstract}

In these lecture notes presented at the 2015 Villa de Leyva Summer School, we give an introduction to superstring theory. We begin by studying the particle and superparticle in order to get a better understanding on the superstring side. Afterwards, we review the pure spinor formalism and end by computing the scattering amplitude for three gravitons at tree-level.

\end{abstract}

\newpage

\tableofcontents

\section{Introduction}

For more than a decade a manifestly super-Poincar\'e covariant formulation for the superstring,
known as the pure spinor formalism \cite{Howe:1991mf,BerkovitsCQS}, has shown to be a powerful framework in two branches.
The first one is the computation of scattering amplitudes and the second one is the quantization of
the superstring in curved backgrounds which can include Ramond-Ramond flux. The strength of
the pure spinor formalism resides precisely in the fact that it can be quantized in a manifestly superPoincar´e
manner, so this covariance is not lost neither in the scattering amplitudes computation
nor in the quantization of the superstring in curved backgrounds.

One key ingredient in this formalism is a bosonic ghost $\l^\a$, constrained to satisfy Cartan’s pure
spinor condition in 10 space-time dimensions \cite{Cartan}. The prescription for computing multiloop   
amplitudes was given in \cite{nathan minimal pure spinor}, where as in the RNS formalism, it was necessary to introduce picture
changing operators (PCO's) in order to absorb the zero-modes of the pure spinor variables. Up to
two-loops, various amplitudes were computed in \cite{two_loop}. Later on, by introducing a set of
non-minimal variables $\lb_\a$ and $r_\a$, an equivalent prescription for computing scattering amplitudes
was formulated in \cite{nathan topological} and \cite{nathan nikita multiloops}. This last superstring description is known as the ``non-minimal"
pure spinor formalism, in order to distinguish it from the former “minimal” pure spinor formalism.
With the non-minimal formalism, also were computed scattering amplitudes up to three-loops \cite{Berkovits:2006bk,Gomez:2013sla}. Because of its topological nature, in the non-minimal version it is not necessary to introduce
PCO's. Nevertheless, it is necessary to use a regulator. The drawback of having to introduce this
regulator appears beyond three-loops, since it gets more complicated due to the divergences coming
from the poles contribution of the b-ghost \cite{Grassi:2009fe}.

In this short note we give an introduction to superstring theory in the pure spinor formalism. We are going to start with very general comments about the superparticle in ten dimensions.

\section{Particle and Superparticle}

We begin this note with a brief introduction to the relativistic point particle and superparticle, please review the references \cite{Polchi, Berkovits:2002zk, superpart, Maframaster, Luismaster}.

A relativistic particle is described by a point in a flat space-time\footnote{The notation $(1,D-1)$ means the metric of the space-time is given by $\eta_{\mu\nu}={\rm dig}(-,+,+,\cdots,+)$} $(1,D-1)$, 
whose evolution over time is described by a curve (worldline).

%%%%%%%%%%%%%%%%%%%%%%
\begin{center}
\includegraphics[scale=1.5]{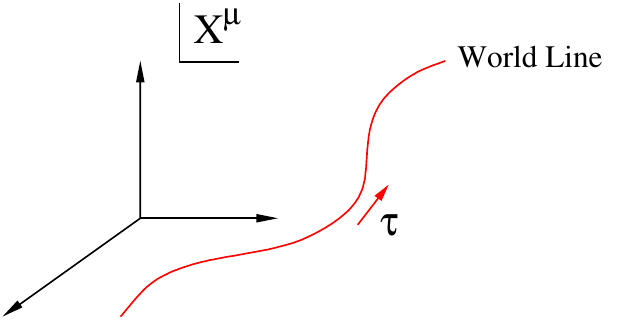}.
\begin{center}
({\bf Fig.1})\\ 
{\small {\rm Point particle evolving over time. The worldline  is parametrized by $\tau$ and the $X^\mu=(X^0,X^i)=(t,\overrightarrow{r})$ are the space-time coordinates. }} 
\end{center}
\end{center}
%%%%%%%%%%%%%%%%%%%%%%%%%%%

The simplest Poincare and $\tau$-reparameterization invariant action is proportional to the worldline length    
\begin{equation}\label{actionone}
S= - M\int ds,
\end{equation}
where $M$ is the mass of the particle. The ``$-$" sign is introduced in order to guarantee that the $S$ functional is going to have a local minimum, i.e. a stable classical trajectory. Let us recall that the space-time induces a metric on the world-line, thus the $ds$ line element is just given by the squared root of the  induced metric 
\begin{equation}\label{differential}
ds=\sqrt{- \eta_{\mu\nu}\dot X^\mu \dot X^\nu}\,\, d\tau.
\end{equation}
Since the worldline is a causal trajectory (see {\bf Fig.2}), i.e. the velocity vector (tangent vector) is a timelike vector
\begin{equation}
 \eta_{\mu\nu}\dot X^\mu \dot X^\nu < 0,
\end{equation}
then one must introduce the ``$-$" sign into the square root so as to obtain a positive number.    

%%%%%%%%%%%%%%%%%%%%%%
\begin{center}
\includegraphics[scale=1]{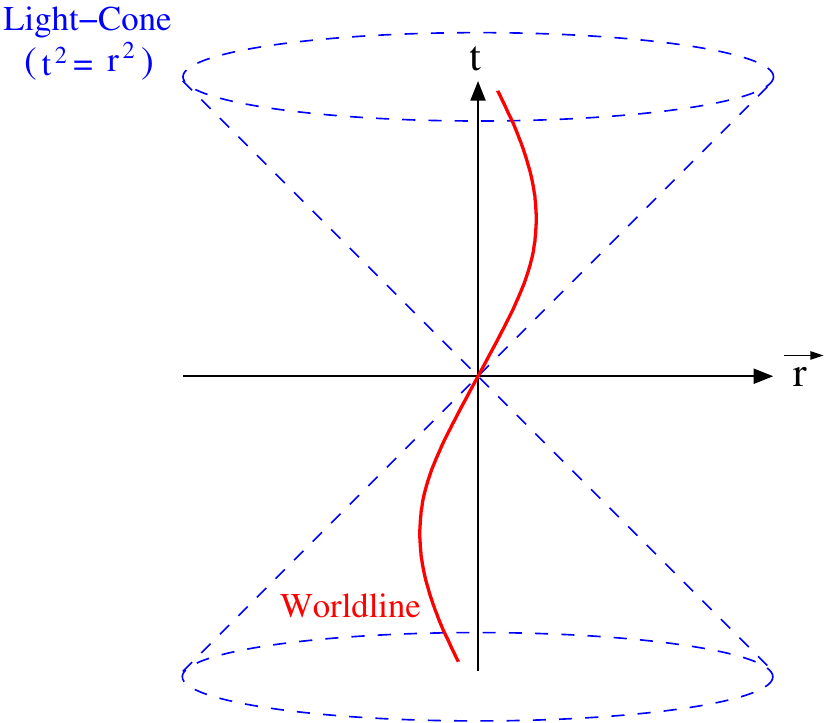}.
\begin{center}
({\bf Fig.2})\\ 
{\small {\rm Causal trajectory. }} 
\end{center}
\end{center}
%%%%%%%%%%%%%%%%%%%%%%%%%%%

Nevertheless, although the action in  \eqref{actionone} seems simple, it is too hard to quantize because we do not know how to perform a path integral with a square root. In addition, this action only describes massive particles, so,  to compute scattering of photons, gluons or gravitons we need to modify it.  
In order to solve these problems, the following first order action can be proposed
\begin{equation}\label{actiontwo}
S=-\int d\tau \left[P^\mu \dot X_\mu + {e\over 2} (P^\mu P_\mu + M^2)\right].  
\end{equation}
This action is classically equivalent to  \eqref{actionone}, i.e. using the $P_\mu$ and $e$ equations of motion. Furthermore, it supports massless particles and its quantization is easier than  \eqref{actionone}. 

Note that \eqref{actiontwo} is invariant, up to total derivative,  by the local (gauge) transformation
\begin{equation}\label{gaugeT}
\delta P^\mu =\xi\,\dot P^\mu ,\qquad \delta X^\mu = \xi \, \dot X^\mu, \qquad \delta e =\xi\, \dot e + \dot \xi\, e\, ,
\end{equation} 
where $\xi=\xi(\tau)$ is a local parameter.
Using this gauge symmetry one can fix the Lagrange multiplier, ``$e$", and perform a BRST quantization. Nevertheless, from \eqref{gaugeT} it is clear that the $e$ field is a 1-form on the worldline, i.e. $e\in H_{\rm dR}^1(C)$, where $C$ is the worldline and $H_{\rm dR}^1(C)$ is the first de-Rham cohomology group over $C$ \cite{harris}.  Therefore, the choice of the gauge fixing depends of the worldline topology. Since here we are not focused on this issue, for more details  see \cite{harris}.

%%%%%%%%%%%%%%%%%%%%%%%%%
\subsection{Brink-Schwarz Superparticle}
%%%%%%%%%%%%%%%%%%%%%%%%%%

As the main topic of this note is to give an introduction to Superstring theory, we will center in a space-time of  ten dimensions. So, we begin with a superparticle in ten dimensions. The main references for this section are \cite{Berkovits:2002zk,Maframaster,Luismaster}.

%%%%%%%%%%%%%%%%%%%%%%%%%%%%%%%
{\it Brink-Schwarz Superparticle}. 
%%%%%%%%%%%%%%%%%%%%%%%%%%%%%%%

The Brink-Schwarz (BS) action for the ten-dimensional (massless) superparticle is given by
\begin{equation}\label{actionBS}
S=\int d\tau \left(\Pi^\mu\, P_\mu + e\, P^\mu\, P_\mu  \right), ~~ {\rm with}~~ \Pi^\mu := \dot X^\mu-{1\over 2} \dot\t^\a \g^\mu_{\a\b} \t^\b, ~~\mu=0,\ldots, 9, ~~ \a,\b=1,\ldots 16,
\end{equation}
where $P^\mu$ is the canonical momentum of $X^\mu$,  $e$ is the Lagrange multiplier to impose the massless condition, $P^2=0$, and $\t^\a$ is a Grassmann or fermionic coordinate\footnote{$\a$ is a (Weyl) spinorial index. When the space-time has $D$-dimensions, where $D$ is an even integer number, then a Weyl spinor has $2^{{D\over 2}-1}$ components.}, i.e. $\t^\a \t^\b=-\t^\b \t^\a$. The Gamma matrices, $\g^\mu_{\a\b}$ and $\g_\mu^{\a\b}$,  are $16\times 16$ symmetric matrices which satisfy the Clifford algebra,  $(\g^\mu)_{\a\b}(\g^\nu)^{\b\r}+(\g^\nu)_{\a\b}(\g^\mu)^{\b\r}=2\eta^{\mu\nu}\de_\a^\r$. In the Weyl representation, $(\g^\mu)_{\a\b}$ and $(\g^\mu)^{\a\b}$ are the
off-diagonal blocks of the $32 \times 32$ Dirac $\G^\mu$ matrices, i.e.
\begin{equation}
\G^\mu=
\left(
\begin{matrix}
0 & (\g^\mu)^{\a\b}\\
(\g^\mu)_{\a\b} & 0
\end{matrix}
\right), \quad {\rm where} ~~~\left\{  \G^\mu, \G^\nu  \right\}=2\eta^{\mu\nu}.
\end{equation}
Besides being invariant by reparameterization
\begin{equation}
\delta P^\mu =\xi\,\dot P^\mu ,\qquad \delta X^\mu = \xi \, \dot X^\mu,\qquad \delta \t^\a = \xi \, \dot \t^\a, \qquad \delta e =\xi\, \dot e + \dot \xi\, e\, ,
\end{equation}
the BS action is invariant under the global transformation
\begin{equation}\label{susy}
\delta \t^\a =  \e^\a, \qquad \delta X^\mu ={1\over 2} \t^\a \g^\mu_{\a\b} \e^\b,\qquad \delta P^\mu=\delta e =0,
\end{equation}
where $\e^\a$ is a constant Grassmann parameter. Using Noether's theorem, this global symmetry is generated by the charge
\begin{equation}
q_\a := p_\a - {1\over 2}\g^\mu_{\a\b} \t^\b P_\mu,
\end{equation}
where 
\begin{equation}\label{constraint}
p_\a:={\p L \over \p \dot\t^\a}=- {1\over 2}\g^\mu_{\a\b} \t^\b P_\mu, 
\end{equation}
is the canonical momentum of $\t^\a$, namely\footnote{${\rm PB}$ means Poisson bracket. Let us  remember that when the two variables are Grassmann numbers, then this commutator becomes an anti-commutator, ``$- ~~\rightarrow~~+ $". }  $\{ p_\b, \t^\a \}_{\rm PB}=-i\de^\a_\b$. It is simple to check 
\begin{equation}
\{ q_\a, q_\b\} = i \g^\mu_{\a\b}P_\mu.
\end{equation}
The charge $q_\a$ is known as the {\it supercharge} and the transformations in \eqref{susy} are the {\it supersymmetry transformations}.  

The BS action is also invariant under the local  transformation
\begin{equation}\label{kappa}
\delta \t^\a =  P^\mu\g_\mu^{\a\b}\kappa_\b, \qquad \delta X^\mu =-{1\over 2} \t^\a \g^\mu_{\a\b} \de\t^\b,\qquad \delta P^\mu=0,  \qquad \delta e=\dot\t^\a \kappa_\a,
\end{equation}
where $\kappa_\a = \kappa_\a(\tau)$ is a local Grassmann parameter. This local symmetry is known as the {\it Kappa symmetry}.  This symmetry is going to be used to perform the light-cone gauge.

From the canonical momentum $p_\a$ obtained in \eqref{constraint}, we obtain a constraint system  given by the conditions
\begin{equation}
d_\a:=p_\a+ {1\over 2}\g^\mu_{\a\b} \t^\b P_\mu=0.
\end{equation}
The algebra of these constraints is given by
\begin{equation}\label{dd}
\{d_\a, d_\b  \}_{\rm PB}=-i\g^\mu_{\a\b} P_\mu.
\end{equation}
Because, $P^2=0$, then one has $8$ first-class constraints and eight second-class constraints.  To  see this we choose a frame where, $P^\mu=(E,0,\ldots, E)$, and later we define the light-cone coordinates and $\g-$matrices as
\begin{equation}\label{frame}
X^{\pm}={1\over \sqrt{2}}(X^0\pm X^9),\qquad P^{\pm}={1\over \sqrt{2}}(P^0\pm P^9),\qquad \g^{\pm}={1\over \sqrt{2}}(\g^0\pm \g^9).
\end{equation}
Since $P^-=0$ and $P^j=0$ for $j=1$ to 8 in this frame,  the algebra in \eqref{dd} becomes
\begin{equation}\label{ddS}
\{d_\a, d_\b  \}_{\rm PB}=-i\g^-_{\a\b} P^+ \propto 
\left(
\begin{matrix}
\mathbb{1}_{8\times 8} & \mathbb{0}_{8\times 8}\\
\mathbb{0}_{8\times 8} & \mathbb{0}_{8\times 8}
\end{matrix}
\right).
\end{equation}

%%%%%%%%%%%%%%%%%%%%%%%%%%%%%%%%
{\it  Gauge Fixing}.
%%%%%%%%%%%%%%%%%%%%%%%%%%%%%%%

Let us recall that to quantize a theory with  second-class constraints the Poisson bracket must be replaced by the Dirac bracket, which is defined as
\begin{equation}\label{DB}
\{A,B\}_{\rm DB}:=\{A,B\}_{\rm PB}-\{A,\phi_i\}_{\rm PB}\, C^{-1}_{ij}\,\{\phi_j,B\}_{\rm PB},
\end{equation} 
where $\phi_i$'s are the second-class constraints and  $C^{-1}_{ij}$ is the inverse matrix of the second-class constraints algebra, $C_{ij}:=\{\phi_i,\phi_j\}_{\rm PB}$.  

For the BS superparticle it is not possible to separate, in a Lorentz covariant way, the first and second-class  constraints, in order to obtain the $C_{ij}$ matrix. However, as it was shown in \eqref{ddS}, there is a frame where the first and second-class  constraints are disjoint, which is known as the light-cone gauge.

To be more precise,  the light-cone gauge consists in choosing a $\t^\a$ field such that $(\g^+\t)_\a=0$, which is possible by the Kappa symmetry.  Since $P^-=0$ and $P^i=0$, $i=1,...,8$, on the frame $P^\mu=(E,0,...,E)$, one can fix $\k_\b={1\over 2 P^+}(\g^+\t)_\b$. Using the $\kappa$ transformation given in \eqref{kappa}, it is straightforward to check
\begin{equation}
\t^{\prime\,\a} = \t^\a + \de \t^\a = -{1\over 2}(\g^+\g^-\t)^\a,
\end{equation}
where we have used, $\{\g^+,\g ^-\}=-1$. So, it is clear that  $(\g^+\t^\prime)_\a=0$. In this gauge, the BS action becomes
\begin{align}\label{actionLCG}
S&=\int d\tau  \left(\Pi^\mu\, P_\mu + e\, P^\mu\, P_\mu  \right)\nonumber\\
& = \int d\tau  \left[\dot X^\mu\, P_\mu -{1\over 2} (-\dot \t \g^+ \t P^- -\dot \t \g^- \t P^+ +\dot \t \g^i \t P^i )+e\, P^\mu\, P_\mu  \right]\nonumber\\
& = \int d\tau  \left(\dot X^\mu\, P_\mu -{1\over 2} \dot S_a S_a +e\, P^\mu\, P_\mu  \right), ~~ a=1,...,8,
\end{align}
where we have utilized $\dot \t \g^+ \t=\dot \t \g^i \t=0$ and defined $S^a=2^{1/4}\sqrt{P^+}\t^a$.  It is useful to remember that a  Weyl spinor  in a $10$-dimensional  space-time can be decomposed in a Weyl and anti-Weyl spinor in an eight-dimensional  space-time, namely
\begin{equation}
\t^\a=
\left(
\begin{matrix}
\t^a\\
\t^{\dot a}
\end{matrix}
\right),\qquad a, \dot a= 1,2,...,8.
\end{equation}
In addition, there is a representation where the $\g^-_{\a\b}$ matrix looks
\begin{equation}
\g^-_{\a\b}=-\sqrt{2}
\left(
\begin{matrix}
\mathbb{1}_{8\times 8} & \mathbb{0}_{8\times 8}\\
\mathbb{0}_{8\times 8} & \mathbb{0}_{8\times 8}
\end{matrix}
\right),
\end{equation}
hence\footnote{The $8$-dimensional  space-time spinor metric is just the identity, $S_a=\de_{ab} S^b$.} ${1\over 2}\dot \t \g^- \t P^+=-{1\over 2} \dot S_a S_a$.

The BS action in the light-cone gauge is more friendly than the original one, but we have lost the Lorentz covariance since the action has eight-dimensional space-time spinor fields.

%%%%%%%%%%%%%%%%%%%%%%%%%%%%%%%%
{\it  Quantization}.
%%%%%%%%%%%%%%%%%%%%%%%%%%%%%%%

From the BS action in \eqref{actionLCG}, the canonical momentum of $S_a$, i.e. $\{p_a, S_b \}_{\rm PB}=-\de_{ab}$, is given by 
\begin{equation}
p_a:=\frac{\p L}{\p \dot S^a}=-{1\over 2}S_a,
\end{equation}
therefore there are  eight constraints,  $d_a=p_a+{1\over 2}S_a =0$. The algebra of these constraints is straightforward to compute
\begin{equation}
\{d_a , d_b \}_{\rm PB}=-\de_{ab},
\end{equation}
which implies that these constraints are of second-class. Thus, using Dirac's method (see \eqref{DB}) we get the anti-commutator
\begin{align}\label{Salgebra}
\{ S_a, S_b\}_{\rm DB}&=\{ S_a, S_b\}_{\rm PB} - \{ S_a, d_c\}_{\rm PB}\,\{ d^c, d^e\}_{\rm PB}^{-1}\,\{ d_e, S_b\}_{\rm PB}\nonumber\\
&= 0 - (-\de_{ac})(-\de_{ce})(-\de_{eb})\\
&= \de_{ab}\nonumber,
\end{align}
which is the Clifford algebra. A representation of this algebra gives us the quantum states of the theory. 

In order to build a representation of \eqref{Salgebra}, it is convenient to keep to mind the $SO(8)$ Pauli matrices\footnote{$SO(8)$ means Special orthogonal group in $8$ dimensions space-time.}, which satisfy
\begin{equation}
\s^i_{a \dot a}\s^j_{\dot a b}+\s^j_{a \dot a}\s^i_{\dot a b}=2\de_{ab}\de^{ij}, \qquad i,j,a,\dot a,b= 1,...,8,
\end{equation}
where $i,j$ are vector indices (space-time) and $a,b,\dot a$ are spinor indices\footnote{The $a$ label is known as a chiral index and $\dot a$ as an anti-chiral.}. Following the Pauli matrices properties, we can represent the algebra in \eqref{Salgebra} using the definitions  
\begin{align}
S_a |\dot a\rangle &=  \frac{1}{\sqrt{2}}\s_{a\dot a}^j  |j\rangle, \\
S_a |i \rangle &=  \frac{1}{\sqrt{2}}\s_{a\dot b}^i  |\dot b\rangle.
\end{align}
Clearly, $\{S_a,S_b\}|\dot a \rangle = \delta_{ab}|\dot a\rangle$ and $\{S_a,S_b\}| i \rangle = \delta_{ab}| i\rangle$, therefore the physical spectrum is a $SO(8)$ vector, given by  $|i\rangle$, and a $SO(8)$ anti-chiral spinor, given by $| \dot a\rangle$, which are massless  by the equation of motion, $P^2=0$. This  is the same spectrum of $D=10$ Super Yang-Mills (SYM), eight degree of freedom (d.o.f) for the gluon and eight d.o.f for the gluino. 

\subsection{Pure Spinor Superparticle}

This section is based on the references \cite{Berkovits:2002zk,superpart}.

As it was shown above, the BS action is read as
\begin{align}\label{BSg}
S=\int d\tau  \left(\dot X^\mu\, P_\mu -{1\over 2} \dot S_a S_a +e\, P^\mu\, P_\mu  \right),
\end{align}
in the light-cone gauge. Nevertheless, we can think that this action comes from a bigger theory, different from the one given in \eqref{actionBS}, such that after fixing the symmetries one obtains \eqref{BSg}. 

Let us consider the following action
\begin{equation}\label{PSaction}
S=\int d\tau  \left(\dot X^\mu\, P_\mu -{1\over 2} \dot S_a S_a +e\, P^\mu\, P_\mu +\dot{\t}^\a p_\a + f^\a \hat d_\a \right),
\end{equation}
where $(\t^a,p_a)$ are independent fermionic fields\footnote{The $\t^\a$ field is not related with $S_a$ as in \eqref{actionLCG}.}, $f^\a$ is a fermionic Lagrange multiplier and $\hat d_\a$ are the fermionic constraints\footnote{It is useful to recall the notation
\begin{equation}
(\g^+)^{\b\r}S_\r = \sqrt{2}
\left(
\begin{matrix}
\mathbb{1}_{8\times 8} & \mathbb{0}_{8\times 8}\\
\mathbb{0}_{8\times 8} & \mathbb{0}_{8\times 8}
\end{matrix}
\right)
\left(
\begin{matrix}
S_a\\
0
\end{matrix}
\right)=(\g^+)^{\b a}S_a
\end{equation}.}
\begin{equation}
\hat d_\a:=d_\a+{1\over \sqrt{P^+}}P_\mu (\g^\mu \g^+ S)_\a,\qquad {\rm with}\quad d_\a:=p_\a+{1\over 2} P_\mu(\g^\mu \t)_\a.
\end{equation}
From the algebra $\{S_a,S_b \}=i\de_{ab}$ and $\{ d_\a, d_\b\}=-iP_\mu \g^\mu_{\a\b}$, it is not hard to check
\begin{equation}
\{ \hat d_\a, \hat d_\b \}= -\frac{i}{2P^+}P^2 (\g^+)_{\a\b},
\end{equation}
where the identities, $(\g^+)^{\de a}  (\g^+)^{\s a}= \sqrt{2}\,(\g^+)^{\de \s}$ and $\{ \g^\mu, \g^\nu  \}_\a ^\b=2 \eta^{\mu\nu}\de_\a^\b$, have been used.
Clearly, the  $\hat d_\a$'s are first-class constraints, which generate a gauge symmetry.  Using this gauge symmetry one can fix, $\t^\a=0$, and so \eqref{PSaction} becomes the BS action. But, the idea is to use the BRST method to quantize this new action (for details of the BRST quantization in superstring theory one can review the references \cite{Polchi}).

From the BRST method, we know that for each gauge symmetry there are ghost and anti-ghost fields with inverse statistics. For example, using the reparametrization gauge symmetry we can fix $e=1/2$,  so
\begin{equation}
\begin{matrix}
{\rm Gauge \,\, Fixing} \qquad & \qquad{\rm Fermionic- ( ghost, \,antighost)}\quad & \qquad {\rm First-class\,\,constraint}\\
e={1 \over 2} \qquad &  \qquad (c, b)\qquad & \qquad P^2=0.
\end{matrix}
\end{equation} 
So, using the gauge symmetry generated by the first-class constraints, $\hat d_\a\approx 0$, we can fix
\begin{equation}
\begin{matrix}
{\rm Gauge\,\, Fixing} \qquad & \qquad{\rm Bosonic - ( ghost, \,antighost)}\quad & \qquad {\rm First-class\,\,constraint}\\
f^\a=0 \qquad &  \qquad (\hat \l^\a, \hat \o_\a)\qquad & \qquad \hat d_\a=0,
\end{matrix}
\end{equation} 
and the action in \eqref{PSaction} becomes
\begin{equation}\label{actionGF}
S=\int d\tau  \left(\dot X^\mu\, P_\mu -{1\over 2} \dot S_a S_a -{1\over 2}\, P^\mu\, P_\mu +\dot{\t}^\a p_\a + \dot c \,b  + \dot{\hat \l}^\a \hat \o_\a \right).
\end{equation}
After fixing the local symmetries and introducing the ghost fields, the gauge symmetries turn into global symmetries, thus using the Noether's procedure one can obtain the conserved charge. That charge is known as the {\it BRST charge}, which is denoted by $Q$,  and in general it can be written as the ghost field times its corresponding constraint (it is a fermionic charge).  In addition, that charge must be nilpotent, i.e. $\{Q,Q\}=Q^2=0$. Therefore, following those ideas one may suspect that the charge should have the form
\begin{equation}
\hat Q=\hat \l^\a \hat d_\a + c P^2,
\end{equation}
but this charge is not nilpotent, $\hat Q^2 =-\frac{i}{2P^+}P^2 (\hat \l\g^+ \hat \l)$. In order to realize a nilpotent BRST charge we must add the term 
\begin{equation}\label{qhat}
\hat Q=\hat \l^\a \hat d_\a + c P^2+ \frac{i}{4P^+}b (\hat \l\g^+ \hat \l),
\end{equation}
which, in fact, arises naturally from the Noether's method.

%%%%%%%%%%%%%%%%%%%%%%%%
{\it Pure Spinor Condition}
%%%%%%%%%%%%%%%%%%%%%%%%

Since the BRST charge is nilpotent, $Q^2=0$, then one can wonder  about its cohomology \cite{harris}, i.e. the coset space defined as 
\begin{equation}
H(Q):= {\rm Ker} Q / {\Im} Q
\end{equation}
where 
\begin{equation}
{\rm Ker} Q:=  \{ \Psi \in {\cal C}^\infty :  Q\Psi =0  \}, \qquad {\rm Im} Q:=  \{ \Psi \in {\cal C}^\infty :  \Psi = Q \Omega  \}.
\end{equation}
Clearly, ${\Im} Q \subset {\rm Ker} Q $. 

In the BRST language, the  physical states are defined as the states which are in the BRST cohomology, i.e
\begin{equation}
H(Q) = \{ \rm Physical ~~ states \}.
\end{equation}
So, to compute the physical states of the action in \eqref{actionGF}, we must find the $\hat Q$ cohomology of the operator in \eqref{qhat}. But, in addition to being a complicated operator,  it is not Lorentz covariant. In \cite{Berkovits:2002zk}, it was shown that the $\hat Q-$cohomology is actually equivalent to the Cohomology of the simple operator
\begin{equation}\label{QL}
Q=\l^\a d_\a, 
\end{equation}
which is independent of $\{S_a,c\}$. Thus, the action in \eqref{actionGF} can be modified to the new and simpler action
\begin{equation}\label{psaction1}
S^{\rm PS}=\int d\tau  \left(\dot X^\mu\, P_\mu -{1\over 2}\, P^\mu\, P_\mu +\dot{\t}^\a p_\a + \dot \l^\a  \o_\a \right).
\end{equation}
As $d_\a$ is not a really first class constraint, $\{d_a,d_b \} = -i P_\mu \g_{\a\b}^\mu $, the BRST charge in  \eqref{QL} is nilpotent if and only if the $\l^\a$ field satisfies the condition
\begin{equation}
\frac{1}{2}Q^2 = \{Q,Q \} = -i P_\mu (\l \g^\mu \l ) ~\Rightarrow~ (\l\g^\mu\l)=0 , ~~ \mu=0,\ldots, 9.
\end{equation}
This condition is known as the pure spinor condition for spinors in ten dimensions. This condition implies that  $\l^\a$ is a complex spinor. For example, let us consider $\mu=0$, i.e.
\begin{equation}
(\l \g^0\l)= -[(\l^1)^2 +(\l^1)^2+\cdots + (\l^{16})^2] =0,
\end{equation}
thus, in order to obtain a non trivial solutions $\l^\a$ must be a complex spinor\footnote{We have used a representation of the Dirac matrices where
\begin{equation}
\g^0_{\a\b}=-
\left(
\begin{matrix}
\mathbb{1}_{8\times 8} & \mathbb{0}_{8\times 8}\\
\mathbb{0}_{8\times 8} & \mathbb{1}_{8\times 8}
\end{matrix}
\right).
\end{equation}}.

In addition, the pure spinor action given in \eqref{psaction1} is invariant under the global transformation
\begin{equation}
\l^\a ~\rightarrow~ e^{iz}\l^\a, \quad  \omega_\a ~\rightarrow~ e^{-iz}\omega_\a.
\end{equation}
By  Noether's procedure the conserved charge is 
\begin{equation}
J:=\l^\a w_\a,
\end{equation}
which is know as the {\it ghost  number}. Clearly $\l^\a$ and $Q$ have ghost number 1 and $\o_\a$ has ghost number $-1$.

{\it Quantization}

In order to find the $Q-$cohomology, it is useful to write the $d_\a$ constraint as an operator. From the canonical momentum representation, $p_\a \rightarrow \frac{\p}{\p\t^\a}$ and $P_\mu \rightarrow \frac{\p}{\p X^\mu}$, we map the $d_\a$ constraint to the operator
\begin{equation}
d_\a = p_\a +{1\over 2} P_\mu(\g^\mu \t)_\a ~~\rightarrow ~~ D_\a=  \frac{\p}{\p\t^\a} +{1\over 2} (\g^\mu \t)_\a \frac{\p}{\p X^\mu}.
\end{equation}
The $D_\a$ operator is known as the super-covariant derivative, and its algebra is just given by $\{ D_\a, D_\b\} = -i\g_{\a\b}^\mu \frac{\p}{\p X^\mu}$. 

Now, we write the most  general super-Poincar\'e covariant wavefunction that can be constructed from $(X^\mu, \t^\a, \l^\a)$
\begin{equation}\label{spf}
\Psi(X ,\t, \l) = C(X, \t) + \l^\a A_\a (X, \t) + (\l \g^{\mu_1,...,\mu_5}\l) A^*_{\mu_1,...,\mu_5}(X, \t) + \l^\a \l^\b \l^\g C^*_{\a\b\g}(X, \t)+\cdots,
\end{equation}
where we have expanded around the bosonic variable, $\l^\a$. The terms in  $\cdots$ include superfields with more than three powers of $\l^\a$(ghost-number greater than three), which  are in the trivial cohomology.

For example, 
$Q\Psi = -i\l^\a D_\a C -i \l^\a \l^\b D_\a A_\b + ...$,
so $Q\Psi=0$ implies that $A_\a(x,\t)$ satisfies the equation of motion
$\l^\a \l^\b D_\a A_\b=0$. But since 
$\l^\a \l^\b$ are pure spinors (see appendix \ref{PSappendix}),  then they are proportional to $(\l\g^{\mu_1\mu_2\mu_3\mu_4\mu_5}\l) \g_{\mu_1\mu_2\mu_3\mu_4\mu_5}^{\a\b}$, 
this implies
that $D\g^{\mu_1\mu_2\mu_3\mu_4\mu_5}A=0$, which is the linearized version of the
super-Yang-Mills
equation of motion. Furthermore, if one defines the
gauge parameter by $\Omega= i\Lambda + \l^\a \omega_\a + ...$, the
gauge transformation $\de\Psi = Q\Omega$ implies $\de A_\a = D_\a \Lambda$
which is the linearized super-Yang-Mills gauge transformation.

So, $A_\a(X,\t)$ contains 
the on-shell super-Yang-Mills gluon and
gluino, $a_\mu(X)$ and $\chi^\a(X)$, which satisfy the linearized equations
of motion and gauge invariances
$$
\p^\mu \p_{[\mu} a_{\nu]} = \g^\mu_{\a\b} \p_\mu\chi^\b =0,\quad \de a_\mu = 
\p_\mu s.
$$
Since gauge invariances of 
antifields correspond to equations of motion of fields, one
expects to have antifields
$a^{*\mu}(x)$ and $\chi^*_\a(x)$ in the cohomology of $Q$
which satisfy
the linearized equations of motion and
gauge invariances
\begin{equation}\label{afeom}
\p_\mu a^{*\mu}=0,\quad \de a^{*\mu} = \p_\nu (\p^\nu s^\mu - \p^\mu s^\nu),\quad
\de\chi_\a^* = \g^\mu_{\a\b}\p_\mu \kappa^\b,
\end{equation}
where $s^\mu $ and $\kappa^\b$ are gauge parameters.
Indeed, these antifields $a^{*\mu}$ and $\chi^*_\a$ appear
in components of the
ghost-number $+2$ superfield $A^*_{\mu_1\mu_2\mu_3\mu_4\mu_5}$ of \eqref{spf}.
Using $Q\Psi=0$ and $\de\Psi = Q\Omega$, $A^*_{\mu_1\mu_2\mu_3\mu_4\mu_5}$
satisfies the linearized
equation of motion $\l^\a(\l\g^{\mu_1\mu_2\mu_3\mu_4\mu_5}\l) D_\a A^*_{\mu_1\mu_2\mu_3\mu_4\mu_5}=0$
with the linearized
gauge invariance $\de A^*_{\mu_1\mu_2\mu_3\mu_4\mu_5} = \g_{\mu_1\mu_2\mu_3\mu_4\mu_5}^{\a\b} D_\a \omega_\b$.
Expanding $\omega_\a$ and $A^*_{\mu_1\mu_2\mu_3\mu_4\mu_5}$ in components, one learns
that 
$A^*_{\mu_1\mu_2\mu_3\mu_4\mu_5}$ can be gauged to the form
\begin{equation}\label{astar}
A^*_{\mu_1\mu_2\mu_3\mu_4\mu_5} = (\t\g_{[\mu_1\mu_2\mu_3}\t)(\t\g_{\mu_4\mu_5]})^\a \chi^*_\a(x) +
(\t\g_{[\mu_1\mu_2\mu_3}\t)(\t\g_{\mu_4\mu_5]s}\t) a^{*s}(x) + ...
\end{equation}
where $\chi^*_\a$ and $a^{*s}$ satisfy the equations of motion 
and residual gauge invariances of \eqref{afeom}, and $...$ involves
terms higher order in $\t^\a$ which depend on derivatives of
$\chi^*_\a$ and $a^{*s}$.

In addition to these fields and antifields, one also expects
to find the Yang-Mills ghost $c(X)$ and antighost $c^*(X)$
in the cohomology of $Q$. The ghost $c(x)$ is found 
in the $\t=0$ component of the ghost-number zero
superfield, $C(X,\t) = c(X) + ...$, and the antighost
$c^*(x)$ is found in the $(\t)^5$ component of the
ghost-number $+3$ superfield,
$C^*_{\a\b\g}(X,\t)= ... + c^*(X) (\g^{\mu_1}\t)_\a (\g^{\mu_2}\t)_\b
(\g^{\mu_3}\t)_\g (\t\g_{\mu_1\mu_2\mu_3}\t) + ... .$ It was proven in \cite{superpart}
that the above
states are the only states in the cohomology of $Q$ and therefore,
although $\Psi$ of \eqref{spf} contains superfields of arbitrarily high
ghost number, only superfields with ghost-number between zero and three
contain states in the cohomology of $Q$.

The linearized equations of motion and gauge invariances $Q\Psi=0$
and $\de\Psi =Q\Omega$ are easily generalized to the non-linear
equations of motion and gauge invariances
\begin{equation}\label{nonls}
Q\Psi + g\Psi \Psi =0,\quad \de\Psi = Q\Omega + g[\Psi,\Omega],
\end{equation}
where $\Psi$ and $\Omega$ transform in the adjoint representation of
the gauge group. For the superfield $A_\a(X,\t)$, \eqref{nonls} implies
the super-Yang-Mills equations of motion and gauge transformations.
Furthermore, the equation of motion and gauge transformation of \eqref{nonls}
can be obtained from the spacetime action\footnote{This spacetime action was first proposed by Edward Witten \cite{stringfield}.}
\begin{equation}\label{yma}
{\cal S}={\rm Tr}\,\,\int d^{10}X\,\, \langle {1\over 2}\Psi Q \Psi + 
{g\over 3}\Psi\Psi\Psi\rangle,
\end{equation}
using the normalization (measure) definition\footnote{This definition will become clear in the following section.} that 
\begin{equation}\label{norm}
\langle (\l\g^{\mu_1}\t)(\l\g^{\mu_2}\t)(\l\g^{\mu_3}\t)(\t\g_{\mu_1\mu_2\mu_3}\t) \rangle =1.
\end{equation}
Although \eqref{norm} may seem strange, it is the only one scalar in the $Q-$cohomology with ghost number three. This measure becomes important in the  superstring scattering amplitudes  context.  After expressing \eqref{yma} in terms of component fields
and integrating out auxiliary fields, it is possible to show
that \eqref{yma} reduces to the standard Batalin-Vilkovisky action for
super-Yang-Mills,
\begin{align}\label{bvym}
{\cal S} =&{\rm  Tr}\int d^{10}X ({1\over 4} f_{\mu\nu} f^{\mu\nu}
+\chi^\a \g^\mu_{\a\b} (\p_\mu\chi^\b +ig[a_\mu,\chi^\b])\\
&+ i a^{*\mu}(\p_\mu c + ig[a_\mu,c]) -g \chi^*_\a \{\chi^\a,c\} -g c c c^*).
\end{align}

%%%%%%%%%%%%%%%%%%%%%%%%
\section{Pure Spinor Superstring}
%%%%%%%%%%%%%%%%%%%%%%%%

In this section we give an introduction to superstring theory using the pure spinor formalism.  Our main objective is to compute, explicitly, the scattering amplitude of  gravitons for three point at tree-level.  This section is based from the references \cite{BerkovitsCQS,Berkovits:2002zk,Mafra:2009wq, dhokerperturbation}.

%%%%%%%%%%%%%%%%%%%%%%%%%
\subsection{General Issues}
%%%%%%%%%%%%%%%%%%%%%%%%%%

From the superparticle pure spinor action found in \eqref{PSaction}, one may integrate out the $P^\mu$ field, so the pure spinor superparticle action becomes  
\begin{equation}\label{psaction}
S^{\rm PS}=\int d\tau  \left(\frac{1}{2}\dot X^\mu\,\dot X_\mu +\dot{\t}^\a p_\a +  \dot\l^\a  \o_\a \right),
\end{equation}
and the BRST charge stays the same.

The most natural and simplest generalization from superparticle to superstring is just to consider a surface instead of worldline curve, i.e.
\begin{align}
(\tau)  ~&\rightarrow ~  (z,\bar z),\nonumber\\
 \{ X(\tau), \t(\tau),p(\tau),\l(\tau), \o(\tau) \}~&\rightarrow ~
\{ X(z,\bar z), \t(z,\bar z),p(z,\bar z),\l(z,\bar z), \o(z,\bar z) \},
\end{align}
and the pure spinor superstring action becomes
\begin{equation}\label{actionSPS}
S^{\rm PS}={1\over 2\pi \a^\prime}\int d^2z  \left(\frac{1}{2}\p X^\mu\,\pb X_\mu +p_\a \pb{\t}^\a  +    \o_\a \pb \l^\a  + \hat p_\a \p{\hat \t}^\a  +    \hat\o_\a \p \hat\l^\a \right),
\end{equation}
where we have denoted $d^2z = dz \, d\bar z$, $\p = \p_{z}, \pb = \p_{\bar z}$ and we introduced the global factor  ${1\over 2\pi \a^\pp}$,  which is the string tension. Furthermore, $\l^\a$ and $\hat\l^\a$ are pure spinors, $(\l\g^\mu\l)=(\hat\l\g^\mu\hat\l)=0$.

Clearly, the complex coordinates parameterize the surface or worldsheet, which is always possible locally. We have also introduced more fields (the hat fields), in order to obtain a real action. Nevertheless, the fermion spinors, $(p_a, \t^\a)$ and $(\hat p_a, \hat\t^\a)$, and the bosonic ones, $(\l^a, \o_\a)$ and $(\hat \l^a, \hat\o_\a)$, may have different chirality, which will define the type of the string. In addition, since the fields are on a surface, they can have different boundary conditions. The boundary conditions depend on whether the surface is open or closed.

For the open string, the boundary conditions are given by\footnote{In the open string, in order to preserve the supersymmetry, it is necessary that the spinors have the same chirality. This string is known as  Type I.  }
\begin{align}\label{openbc}
\p X^\mu &= \pb X^\mu\nonumber\\
\t ^\a (z) &= \hat \t^\a(\bar z)\nonumber\\
p _\a (z) &= \hat p_\a(\bar z),\qquad {\rm when ~~} z=\bar z.\\\
\l ^\a (z) &= \hat \l^\a(\bar z)\nonumber\\
\o _\a (z) &= \hat \o_\a(\bar z)\nonumber
\end{align}
It is useful to remember that the equations of motion of the pure spinor superstring action are 
\begin{align}\label{eofm}
&\p\pb X^\mu=0,\nonumber\\
&\pb\t^\a=\pb p_\a= \pb\l^\a= \pb\o_\a =0,\\
&\p \hat\t^\a=\p \hat p_\a= \p \hat\l^\a= \p \hat\o_\a =0.\nonumber
\end{align}
Therefore, the holomorphic fields, $\{\t^\a,p_\a, \l^\a, \o_\a\}$, are known as the left sector and the antiholomorphic fields, $\{\hat \t^\a, \hat p_\a, \hat\l^\a, \hat \o_\a\}$, are the right sector.  

The boundary conditions of the closed string are just given by the periodicity, for example,
\begin{equation}
\pb X^\mu(z+2\pi) = \pb X^\mu(z) , ~~ \t^\a(z+2\pi) = \t^\a(z),  
~~ \l^\a(z+2\pi)= \l^\a(z), ~~ \ldots
\end{equation}
In the closed string, when the fields, $\{\t^\a,p_\a, \l^\a, \o_\a\}$ and $\{\hat \t^\a, \hat p_\a, \hat\l^\a, \hat\o_\a\}$, have the same chirality, it is called {\it string type} $II B$. When the fields,  $\{\t^\a,p_\a, \l^\a, \o_\a\}$ and $\{\hat \t_\a, \hat p^\a, \hat\l_\a, \hat\o^\a\}$, have the opposite chirality,  then this string  is called {\it string type} $II A$.

The BRST charge looks very similar to the one found in the  superparticle
\begin{equation}
Q:=\int dz \, ( \l^\a d_\a), \qquad  \bar Q:=\int d\bar z \, (\hat \l^\a \hat d_\a).
\end{equation}
We have  now two BRST charges, holomorphic and antiholomorphic, which are independent in the closed string. The $d_\a (\hat d_\a)$ constraint is a little different than the one obtained in superparticle, which is written as
\begin{equation}
d_\a := p_\a -{1\over 2}(\g^\mu \t)_\a \p X_\mu -{1\over 8}
(\g^\mu\t)_\a (\t \g_{\mu}\p\t), 
\end{equation}
and its algebra is, $\{d_\a, d_\b \}=-\g ^\mu_{\a\b} \Pi_\mu$, where $\Pi^\mu = \p X_\mu +{1\over 2}
 (\t \g_{\mu}\p\t) $ is known as the supersymmetric momentum\footnote{The definition of $\hat d_\a$ is, $\hat d_\a := \hat p_\a -{1\over 2}(\g^\mu \hat \t)_\a \pb X_\mu -{1\over 8}
(\g^\mu \hat\t)_\a (\hat \t \g_{\mu}\pb\hat\t) $.}. This constraint arises naturally from the Green-Schwarz action for superstring, but we will not consider it here\footnote{In the rest of the document we only work with the left sector.}.

%%%%%%%%%%%%%%%%%%%%%%%%%%%%
\subsection{Some Symmetries}
%%%%%%%%%%%%%%%%%%%%%%%%%%

It is very useful to remember that in the superparticle case we had gauged the  reparametrization invariance by fixing $e=-1/2$. On the worldline the e-field is interpreted as its metric.  Therefore, on the string side the generalization of the $e$-field is the two dimensional metric, $g_{ab}, a,b=1,2$, but the reparameterization-invariant superstring pure spinor action it is not very well understood.  In addition, the action in \eqref{actionSPS} has the remnant symmetry which is known as conformal symmetry  (holomorphic transformations)
\begin{equation}
z ~~\rightarrow ~~ z^\prime = z^\prime (z), \quad {\rm Holomorphic ~~ transformation.}
\end{equation}
Since the fields, $(X^\mu, \t^\a, \l^\a)$, are scalars on the worldsheet and $(p_\a, \o_\a)$ are $(1,0)$ differential forms then the current conserved is
\begin{equation}
T(z) = -\frac{1}{2}\p X^\mu \p X_\mu - p_\a\p\t^\a + \o_\a \p \l^\a,
\end{equation}
that is known as the holomorphic stress tensor. Its anti-holomorphic counterpart is just given by the fields with hat.

The pure spinor superstring action has also the global symmetries
\begin{align}\label{transformations}
\begin{matrix}
&& {\rm space-time ~~ supersymmetry}  \qquad &\quad&\ {\rm ghost-number}  \\
 & \\
&& \de \l^\a = \de\o_\a =0      \qquad  &\quad  &\de \l^\a = e^{i \a} \l^\a          \\
&&\de X^\mu =   \frac{1}{2}(\e \g^\mu \t)     \qquad  & \quad & \de \o_\a = e^{-i \a} \o_\a       \\
&&\de \t^\a =\e^\a          \qquad    &\quad & \de X^\mu =0           \\
&&\de p_\a = -\frac{1}{2} (\e \g^\mu)_\a\p X_\mu  + \frac{1}{8}(\e\g^\mu \t) (\p\t \g_\mu)_\a                      
\qquad &\quad & \de \t^\a=\de p_\a=0
\end{matrix}
\end{align}
These symmetries give us the charges
\begin{align}
q_\a &= - \int dz \,\left( p_\a + \frac{1}{2} (\e \g^\mu)_\a\p X_\mu  + \frac{1}{24}(\t\g^\mu \p\t) (\t \g_\mu)_\a \right),\qquad {\rm Supercharge} \\                
G &= \int dz J(z) = \int dz \,(\l^\a \o_\a) ,\qquad \qquad \qquad \qquad \qquad \qquad{\rm Ghost-number}.
\end{align}

The Poincare invariance, which can be written as 
\begin{align}
\de X^{\mu} &= \L_{\nu}^\mu X^\nu+a^\mu,\qquad\qquad \qquad \\
\de\t^{\a} &={1 \over 4} \L_{\mu \nu}(\g^{\mu\nu}\t)^\a,\qquad  \de p_{\a} = {1 \over 4} \L_{\mu \nu}(\g^{\mu\nu}p)_\a,\\
\de\l^{\a} &= {1 \over 4} \L_{\mu \nu}(\g^{\mu\nu}\l)^\a,\qquad  \de\o_{\a} = {1 \over 4} \L_{\mu \nu}(\g^{\mu\nu}\o)_\a,
\end{align}
where $\L_{\mu\nu}=-\L_{\nu\mu}$, it is generated by the currents
\begin{align}
P^{\mu} &= \p X^\mu,  \qquad\qquad L^{\mu\nu} = X^\mu \p X^\nu - X^\nu \p X^\mu, \\
\Sigma ^{\mu\nu} &= {1 \over 2} (p \g^{\mu\nu}\t)\\
N^{\mu\nu} &= {1 \over 2} (\o \g^{\mu\nu}\l).
\end{align}

Finally, the pure spinor action in \eqref{actionSPS} has an extra  local  symmetry as a consequence of the pure spinor constraint, $(\l\g^\mu\l)=0$, which is given by
\begin{equation}\label{gaugew}
\de \o_\a = \L_\mu (\g^\mu \l)_\a.
\end{equation}

The pure spinor constraint implies that the number of degrees of freedom of $\l^\a$ is just 11 (see appendix \ref{PSappendix}), in addition, using the local symmetry in \eqref{gaugew} one can fix 5 of the 16 components of $\o_\a$. Hence the number of degrees of freedom of $\l^\a$ and $\o_\b$ is the same, 11.

%%%%%%%%%%%%%%%%%%%%%%%%%%%%
\subsection{OPE's and Anomaly}
%%%%%%%%%%%%%%%%%%%%%%%%%%

In two dimensional theories, particularly in conformal theories, one often has to compute the OPE's among different physical operators. The OPE's give us many information about the theory,  such as the topology of the target space, anomalies, symmetries and  amplitudes. 
For this section one can review  \cite{BerkovitsCQS,nathan minimal pure spinor,betagammasystem,nathan nikita multiloops}.

Roughly speaking, the OPE's are defined just as the correlation function  between operators. In addition, as it is well known from quantum field theory, a correlation function is just  a Green function of some operator. For example, from the pure spinor action, it is simple to see that  the correlation function among $X^\mu$ with itself is just the Green function of Laplace the operator $\p\pb$, namely (on the sphere)
\begin{equation}
\langle X^\mu (z) X_{\nu}(y)\rangle : = X^\mu (z) X^{\nu}(y) = -\frac{\eta ^{\mu\nu}}{2}\,{\rm ln} |z-y |^2 + {\rm reg}\,\, ,
\end{equation}
where $``{\rm reg}"$ meas regular terms in $(z-y)$. In the similar way,  OPE'e among the others fields are
\begin{align}
p_\a(z)\t^\b(y) &= \frac{\de_\a^\b}{z-y} + {\rm reg},\\
 \o_\a(z)\l^\b(y) &= \frac{\de_\a^\b}{z-y} + {{\rm correction\,\, from \,\, the \,\, pure \,\, spinor \, \, condition }\over z-y} + {\rm reg} ,
\end{align}
where the correction from the pure spinor condition is a little complicated and for more details see \cite{BerkovitsCQS}. 

Using the previous fundamental OPE's and applying the  Wick theorem, we can compute the OPE's  among the differents currents. For instance, let us consider the followings two OPE's
\begin{align}
T(z) T(y) &= \frac{2}{(z-y)^2}T(y) + \frac{1}{(z-y)} \p T(y) +{\rm reg} \label{TT}\\
T(z) J (y) & =  \frac{8}{(z-y)^3} + \frac{1}{(z-y)^2} J(y) + \frac{1}{(z-y)} \p J(y)  \label {TJ}.
\end{align} 
The first one means that the pure spinor formalism is free of conformal anomaly. In general, the OPE among the stress tensor with itself   is given by 
\begin{equation}
T(z) T(y) = \frac{c}{2\,(z-y)^4} +\frac{2}{(z-y)^2}T(y) + \frac{1}{(z-y)} \p T(y) +{\rm reg}\label{TTG} .
\end{equation}
The first term is the anomalous term and the $``c"$ constant is known as the central charge. Theories with non-zero central charge have a conformal anomaly, i.e. at quantum level the conformal symmetry in these theories is broken. Furthermore,  since the stress tensor is the generator of the conformal transformation then the quadratic pole of the OPE with $T(z)$ gives the conformal weight, namely how the field transforms  under a conformal transformation.  Clearly, $T(z) = T_{zz}(z)$ has corformal weight 2, to wit $T^\prime(z^\prime) = (\p_{z^\prime}z)^2 T(z)$.

From the second OPE in \eqref{TJ},  one can see that the ghost current has conformal weight 1, as it was expected since that $\l^\a$ is a world-sheet scalar and $\o_\a$ is an holomorphic form,  in addition this current has an anomaly given by the number $8$ in the cubic pole.  As $J(z)$ just depends on the pure spinor and its conjugate momentum then this  anomaly  gives us topological properties of the pure spinor space. Since the pure spinor action is invariant under the ghost number transformation (see \eqref{transformations}), this implies that the anomaly is present in the integration measure of the path integral, i.e.
\begin{equation}
[{\cal D}\l] [{\cal D}\o] ~\rightarrow~ {\rm ghost\,\, number\,\, }8.
\end{equation}

The $\o$ field, which is a differential form of weight $(1,0)$ over the world-sheet,  can be expanded as a linear combination of the eigenfunctions of the operator $\pb$, namely
\begin{equation}
\o_\b = \sum_{i} \o_\b^i\,f_i(z, \bar z) ,\qquad {\rm where ~~} \pb \,f_i(z,\bar z) = \g_i f_i(z, \bar z).
\end{equation}  
Let us recall that on the sphere the only global holomorphic forms are constant functions, so there is no eigenfunction of conformal weight $(1,0)$ with eigenvalue zero, i.e. $\g_i \neq 0$. The eigenfunctions with eigenvalue zero are called the {\it zero modes}, so the $\o_\a$ field does not have zero modes on the sphere and the measure $[{\cal D}\o]$ reads
\begin{equation}
[{\cal D}\o] = \prod_{i=1}[d\o_\b^i], 
\end{equation}
where $[d\o_\b^i]$ is the $\o_\b$ measure  over the phase space $(\l^\a,\o_\b)$. Now,  as the $\l^\a$ is a scalar field of conformal weight $(0,0)$ over the world-sheet, then it can be expanded as a linear combination of the eigenfunctions of the operator $\pb$, i.e.
\begin{equation}
\l^\a = \sum_{i} \l^\a_i\,h_i(z, \bar z) ,\qquad {\rm where ~~} \pb \,h_i(z,\bar z) = \r_i\, h_i(z, \bar z).
\end{equation}  
Since the constant functions $h_0$ are zero modes of conformal weight $(0,0)$, the measure  $[{\cal D}\l]$ becomes
\begin{equation}
[{\cal D}\l] = [d\l^\a_0]\, \prod_{i=1}[d\l^\a_i], 
\end{equation}
where $[d\l^\a_i]$ is the holomorphic measure of the pure spinor space. Therefore the total measure can be written as
\begin{equation}
[{\cal D}\l] [{\cal D}\o] =  [d\l^\a_0]\, \prod_{i=1}[d\l^\a_i]\,[d\o_\b^i].
\end{equation}
Since $\l^\a$ has ghost number $1$ and $\o_\b$ has ghost number $-1$, then the measure $\prod_{i=1}[d\l^\a_i]\,[d\o_\b^i]$ has ghost number 0,  thus we conclude that the ghost number anomaly is just given by the measure of the zero modes
\begin{equation}
[d\l^\a_0] ~\rightarrow~ {\rm ghost\,\, number\,\, }8.
\end{equation}

In order to compute scattering amplitudes, we must build a top holomorphic form, $[d\l^\a_0]$, to wit  an $11-$form, with ghost number 8.   This top holomorphic form can be written in the following covariant way\footnote{It is useful to see the appendix \ref{PSP}.}
\begin{equation}\label{top-form}
[d\l^\a] (\l\g^{\mu_1})_{\a_1} (\l\g^{\mu_2})_{\a_2}(\l\g^{\mu_3})_{\a_3} (\g_{\mu_1\mu_2\mu_3})_{\a_4\a_5} = \e_{\a_1\ldots \a_5\b_1\ldots\b_{11}}d\l^{\b_1}\wedge \cdots \wedge d\l^{\b_{11}},
\end{equation}
where $\e_{\a_1\ldots \a_5\b_1\ldots\b_{11}}$ is the 16-dimensional  totally antisymmetric tensor (Levi-Civita symbol) and we have removed the zero modes subindex ``0". Using the pure spinor constraint and the $\g-$matrices algebra, it is not hard to check that, in fact, the term $(\l\g^{\mu_1})_{\a_1} (\l\g^{\mu_2})_{\a_2}(\l\g^{\mu_3})_{\a_3} (\g_{\mu_1\mu_2\mu_3})_{\a_4\a_5} $ is totally antisymmetric in the spinorial labels. Clearly, the left and right side of the equality in \eqref{top-form} have ghost number 11 and the term on the left hand is the same one which appears in  \eqref{norm}.

\subsection{Massless states}
 
In order to give a prescription to compute scattering amplitudes in the pure spinor formalism, it is needed to introduce the vertex operators, namely to find the BRST Cohomology. This section is going to be brief due to the long computations to check the results, for more details see reviews \cite{BerkovitsCQS, Berkovits:2002zk, Berkovits:2002qx,Mafra:2009wq} 

The physical states in the pure spinor formalism are defined as ghost-number one states in the BRST cohomology of $Q=\int dz (\l^\a d_\a)$. In addition, since we are just interested in massless states then they must have conformal weight zero by the relation $(mass)^2=k^2=\frac{n}{2}$, where $n$ is the conformal weight and $k^\mu$ is the momentum vector. So, the most general massless operator at ghost number zero is
\begin{equation}\label{UvertexO}
V(z)=\l^\a A_\a(X,\t).
\end{equation}
which is known as the unintegrated vertex operator.
From the BRST cohomology condition, 
$Q\, V=0$,  one obtains the constraint \\
$$(\g_{\mu_1\mu_2\mu_3\mu_4\mu_5})^{\a\b}D_\a A_\b=0 ,$$
which is the equation of motion for the spinor potential of super-Yang-Mills.  Furthermore, the gauge transformation   
$$\delta V = Q \,\Omega(X,\t) = \l^\a D_\a \Omega(X,\t),  $$ 
reproduces the usual super-Yang-Mills gauge transformation $\delta A_\a = D_\a\Omega(X,\t) $, where $\Omega(X,\t)$ is a generic scalar superfield. So, the ghost number 1 cohomology of $Q$ for the massless sector
reproduces the desired super-Yang-Mills spectrum.

It is possible to show there is a gauge such that
\begin{equation}
A_\a(X,\t)={1\over 2}a_\mu (X) (\g^\mu\t)_\a-{1\over 3}(\xi(X)\g_\mu\t)(\g^\mu\t)_\a- {1\over 16}\p_{[\mu}a_{\nu]}(\t \g^{\de\mu\nu} \t)(\g_\de\t)_\a+\ldots   \,\, ,
\end{equation}
where $a_\mu(X)=e_\mu \,e^{ik\cdot X}$ and $\xi^\a(X)=\chi^\a \,e^{ik\cdot X}$ are the gluon and gluino fields of the SYM theory and $e_\mu$ and $\chi^\a$ are the polarization vectors and $``[\mu,\nu]"$ is the antisymmetrization of the indices. 

The unintegrated vertex operators are needed to fix the global symmetry over the Riemann surface. For example on the sphere (tree-level amplitude)  the global symmetry group is $PSL(2,\mathbb{C})$, which has three generators. So, in order to fix this global symmetry, one must use three unintegrated vertex operators in the scattering amplitudes prescription, which can be fixed at any point. The others vertex operators in the scattering amplitudes prescription are integrated vertex operators.  The integrated vertex operators, which we will call as $U(z)$, associate with the unintegrated vertex operator $V$ is defined to satisfy
\begin{equation}
Q\,U(z) = \p_z \, V(z).
\end{equation}
Note that , $Q(\int U(z))=0$. From this definition one can check that the integrated vertex operator associated to $V(z)=\l^\a\, A_\a(X,\t)$ is 
\begin{equation}
U(z) = \p_z \t^\a A_\a(X,\t)+\Pi^\mu B_\mu (X,\t)+d_\a W^\a(X,\t)+{1\over 2} N_{\mu\nu}{\cal F}^{\mu\nu}(X,\t)
\end{equation}
where the superfields, $\{B_\mu (X,\t), W^\a(X,\t), {\cal F}^{\mu\nu}(X,\t)\}$, satisfy the constraints 
\begin{align}
&D_\a A_\b + D_\b A_\a - \g_{\a\b}^\mu B_\mu =0,\\
&D_\a B_\mu - \p_\mu A_\a - (\g_\mu)_{\a\b} W^\b =0,\\
&D_\a W^\b - {1\over 4} (\g_{\mu\nu})_{\a}^{\,\,\,\b} {\cal F}^{\mu\nu} =0,\\
& \l^\a \l^\b (\g_{\mu\nu})_{\b}^{\,\,\,\g} D_\a {\cal F}^{\mu\nu}=0,
\end{align}
which imply the super-Maxwell equations of motion.

For the closed string the vertex operators are just the tensorial product of operators from the left and right  sector, to wit
\begin{align}
V_{\rm closed}&= V(z) \otimes \hat V(\bar z)= \l^\a \hat\l^\b A_\a(\t) \otimes \hat A_\b (\hat \t) \, e^{i \,k\cdot X},\\
U_{\rm closed}&= U(z) \otimes \hat U(\bar z),
\end{align}
where the graviton, $g_{\mu \nu}$, is identified with $e_\mu\otimes \hat e_{\nu}$ and the gravitino, $\psi_\mu ^\a$ ($\hat \psi_\mu ^\a$),  with $e_\mu\otimes \hat\chi^\a$ ($\chi^\a \otimes \hat e_\mu$).

\subsection{Tree-Level Scattering Amplitudes}

For more details of this section one can review  \cite{BerkovitsCQS,nathan minimal pure spinor,nathan topological}

In this section we give an example how to compute  scattering amplitudes at tree-level using the pure spinor formalism, in particular for the closed string, i.e. on a sphere. 

In general, the scattering amplitude prescription on a sphere is given by the expression
\begin{equation}\label{prescriptionPS}
{\cal M}_{n}:=\prod_{i=4}^n \int d^2z_i \,\,\left\langle\Big| V(z_1)\,V(z_2)\,V(z_3)\,U(z_4)\,\cdots U(z_n) \Big|^2\right\rangle,
\end{equation}
where the power two is due to left and right sector. The three unintegrated vertex operators fix the $PSL(2,\mathbb{C})$ global symmetry and the  points $\{z_1,z_2,z_3\}$ are arbitrary on the sphere, which often are chosen to be $z_1=1, z_2=0, z_3=\infty$. The triangular bracket, $\langle \cdots \rangle$, means integration by all fields, i.e 
\begin{equation}
\langle\cdots \rangle = \int [{\cal D}X][{\cal D}\l][{\cal D}\o][{\cal D}\t][{\cal D}d]\,\, \cdots,
\end{equation}
where we have replaced the $[{\cal D}p]$ integration  by $[{\cal D}d]$. 

Since $\l^\a$ and $\o_\a$ are complex variables then the integration by these variables must be a contour integral. The contour can be fixed introducing the Cauchy kernel (delta Dirac function), which are known as the picture changing operators.  Nevertheless, in 2005 Berkovits and Nekrasov introduced a new set of fields, the complex conjugate of $(\l^\a, \o_\a )$, i.e. $(\bar \l_\a , \bar\o^\a )$, in order to integrate over the whole pure spinor space. In addition, so as to keep the central charge, $c=0$ (see \eqref{TTG}), two more fermionic fields must be introduced, $(r_\a, s^\a)$,  where $r_\a$  is constrained to satisfy\footnote{Note that the $r_\a$ field can be interpreted as an antiholomorphic form over the pure spinor space, to wit  $r_\a \equiv d\bar\l_\a$.}, $(\bar\l  \g^\mu r)=0$, $\mu=0,...,9$.  The BRST charge is also modified\footnote{Clearly the operator $\int (r_\a \bar \omega^\a )$ can be identify with the Dolbeault operator $d\bar\l_\a \frac{\p}{\p\bar\l_\a}$. So, $\tilde Q$ is an equivariant operator.}
\begin{equation}
Q=\int dz (\l^\a d_\a)\,\,\, \longrightarrow \,\,\, \tilde Q=\int dz (\l^\a d_\a + \bar\omega^\a r_\a),
\end{equation}
but the cohomology of $Q$ and $\tilde Q$ are the same. 

In this new version,  the ghost anomaly is $-3$, i.e.
\begin{equation}
[{\cal D}\t][{\cal D}d][{\cal D}\l][{\cal D}\bar\l][{\cal D}\o][{\cal D}\bar\o][{\cal D}r][{\cal D}s]\,\,\,\longrightarrow\,\,\,{\rm ghost\,\, number}\,-3,
\end{equation}
where the ghost current is given by $J(z)=(\o_\a\l^\a)-(\bar\o^\a\bar\l_\a)$. But, the total integral given in \eqref{prescriptionPS} has ghost number zero, to wit 
\begin{equation}
[{\cal D}\t][{\cal D}d][{\cal D}\l][{\cal D}\bar\l][{\cal D}\o][{\cal D}\bar\o][{\cal D}r][{\cal D}s] V(z_1)V(z_2)V(z_3) U(z_4)\cdots U(z_n) \,\,\,\longrightarrow\,\,\,{\rm ghost\,\, number}\,\, 0.
\end{equation} 

It is not hard to check that the integration, $\int [{\cal D}\t][{\cal D}d] [{\cal D}\l][{\cal D}\bar\l][{\cal D}\o][{\cal D}\bar\o][{\cal D}r][{\cal D}s]\cdots $, is equivalent to  the bracket
\begin{equation}\label{measurePSS}
\int [{\cal D}\t][{\cal D}d] [{\cal D}\l][{\cal D}\bar\l][{\cal D}\o][{\cal D}\bar\o][{\cal D}r][{\cal D}s]\cdots \,\,\,\longrightarrow\,\,\, \left\langle (\l \g^{\mu_1}\t) (\l \g^{\mu_2}\t) (\l \g^{\mu_3}\t) (\t \g_{\mu_1\mu_2\mu_3}\t)  \right\rangle= {\rm C},
\end{equation} 
where ${\rm C}$ is a constant. In general this constant is normalized to be ${\rm C}=1$, so as in \eqref{norm}.

\subsubsection{Three Gravitons at Tree-Level}

In this example we compute a scattering amplitude at tree-level  for three gravitons. This is the simplest case since the integrated vertex operators are not needed\footnote{This section is based on the C. Mafra's master thesis \cite{Maframaster}.}.

The amplitude is given by
\begin{equation}
{\cal M}_3 = \left\langle \Big|    V(z_1)V(z_2)V(z_3)   \Big|^2 \right\rangle,
\end{equation}
with
\begin{equation}\label{expansionV}
V(z_j)={1 \over 2} e^j_\mu (\l\g^\mu \t) e^{i k_j\cdot X} -{1\over 16 }k^j_{\mu} e^j_\nu (\l\g_\rho \t)(\t \g^{\mu\nu\rho}\t)e^{i k_j\cdot X} + ... 
\end{equation}
where $...$ includes terms quartic and higher-order in $\theta$ and we have just considered the bosonic contribution, i.e. the polarization vector $e^j_\mu$, where $j$ is the label of the corresponding particle. 

From the integration given in \eqref{measurePSS}, the only non-zero contributions are those in which there are five $\t'$s. So, following the expansion in \eqref{expansionV}, there are just three possibilities to distribute the $\t$ field, $(1,1,3),(1,3,1),(3,1,1) $.

The first contribution is given by
 \begin{equation}
{\cal M}^1_3 = 
e^1_{\mu_1} e^2_{\mu_2} k^3_{\nu_3} e^3_{\mu_3}
\left\langle 
(\l \g^{\mu_1}\t) (\l \g^{\mu_2}\t) (\l \g_{\rho}\t) (\t \g^{\rho\nu_3\mu_3}\t)  \right\rangle 
\left\langle  e^{i k_1\cdot X(z_1)}e^{i k_2\cdot X(z_2)}e^{i k_3\cdot X(z_3)}   \right\rangle.
\end{equation}
The integration by the $X^\mu$ field is simple and the answer is
\begin{align}
\left\langle  e^{i k_1\cdot X(z_1)}e^{i k_2\cdot X(z_2)}e^{i k_3\cdot X(z_3)}   \right\rangle &= \int [{\cal D}X]e^{-\int d^2z\, \p X\cdot \pb X}  e^{i k_1\cdot X(z_1)}e^{i k_2\cdot X(z_2)}e^{i k_3\cdot X(z_3)}\nonumber \\
& = |z_{12}|^{2 k_1\cdot k_2}|z_{13}|^{2 k_1\cdot k_3}|z_{23}|^{2 k_2\cdot k_3},\label{Xintegration}
\end{align}
where $z_{ij} : = z_i - z_j$. From the on-shell condition, $k_j^2=0$, and the momentum conservation constraint, $k_1^\mu+k_2^\mu+k_3^\mu=0$, it is trivial to check, $k_1\cdot k_2 =k_1\cdot k_3 =k_2\cdot k_3 =0$, therefore 
\begin{align}
\left\langle  e^{i k_1\cdot X(z_1)}e^{i k_2\cdot X(z_2)}e^{i k_3\cdot X(z_3)}   \right\rangle =1 .
\end{align}
Note that we have not introduced the functional determinant, ${\rm det}(\p\pb)$, in \eqref{Xintegration}, the reason is because it will be canceled out by the other functional determinants.

Up to an overall factor, it is not hard to check\footnote{For more details, see appendix of \cite{Berkovits:2006bk}.}
\begin{equation}
\left\langle 
(\l \g^{\mu_1}\t) (\l \g^{\mu_2}\t) (\l \g_{\rho}\t) (\t \g^{\rho\nu_3\mu_3}\t)  \right\rangle
\propto \eta^{\mu_1 \nu_3} \eta^{\mu_2 \mu_3}- \eta^{\mu_1 \mu_3} \eta^{\mu_2 \nu_3}.
\end{equation}
Finally, the contribution $(1,1,3)$ becomes
 \begin{equation}
{\cal M}^1_3 = 
(e^1\cdot k^3)( e^2\cdot  e^3) - (e^2\cdot k^3 )( e^1\cdot  e^3). 
\end{equation}
In a similar way,  the contributions $(1,3,1)$ and $(3,1,1)$ are given by
 \begin{align}
{\cal M}^2_3 &= -
(e^1\cdot k^2)( e^2\cdot  e^3) + (e^3\cdot k^2 )(e^1\cdot  e^2), \\ 
{\cal M}^3_3 &= 
(e^2\cdot k^1)( e^1\cdot  e^3) - (e^3\cdot k^1) ( e^1\cdot  e^2). 
\end{align} 
Therefore, the total amplitude reads
\begin{equation}
{\cal M}_3 = \Big|  {\cal M}^1_3+{\cal M}^2_3+{\cal M}^3_3 \Big|^2
=\Big| 2 ( e^1\cdot  e^2)  (e^3\cdot k^2)+2( e^1\cdot  e^3)  (e^2\cdot k^1) +2( e^2\cdot  e^3)  (e^1\cdot k^3) \Big|^2,
\end{equation}
where we have use the momentum conservation, $k^\mu_1+k^\mu_2+k^\mu_3=0$, and the transversality condition, $e^j\cdot k^j =0$. Up to overall constant,  this is the right result \cite{Polchi}

%%%%%%%%%%%%%%%%%%%%%%%%%%%%%%%%%%%%%%%%%%%%%%%%%%%%%%%%%%%%%%%%%%%
%%%%%%%%%%%%%%%%%%%%%%%%%%%%%%%%%%%%%%%%%%%%%%%%%%%%%%%%%%%%%%%%%%%

\section*{Acknowledgments}

We thank the organizers of the Villa de Leyva Summer School for their hospitality.
We thank S. Mizera, G. Zhang, C. Mafra and H. Ocampo for reading the manuscript. H.G. would like to thank to M. Guillen for discussions about superparticle. H.G. is very grateful to the Perimeter Institute for hospitality during this work. The work of  H.G.  is supported by USC grant DGI-COCEIN-No 935-621115-N22 and the work of N.B. is supported by FAPESP grants 2016/01343-7 and 2014/18634-9 and CNPq grant 300256/94-9.

\appendix

\section{Cartan and Chevalley definitions.}\label{PSappendix}

This appendix is based on the lectures on beta-gamma system  given in  \cite{betagammasystem}.

The $SO(2d)$ pure spinor \ $\l^\a$ is constrained
to satisfy \cite{Cartan}
\begin{equation}
\l^\a (\g^{\mu_1 ...  \mu_j})_{\a\b} \lambda^\b =0 , \qquad  {\rm for} \qquad 0\leq j< d \ ,
\end{equation}
where $\mu=1$ to $2D$, $\a=1$ to $2^{d-1}$, and $\g^{\mu_1 ...\mu_j}_{\a\b}$ is the
antisymmetrized
product of $j$ Pauli matrices, i.e.
\begin{equation}
\g^{\mu_1 ...\mu_j}:=\frac{1}{j!}\g^{[\mu_1}\g^{\mu_2}\ldots \g^{\mu_j]}.
\end{equation} 
This implies that
$\l^\a \l^\b$ can be written as
\begin{equation}
\l^\a \l^\b =
{{1}\over{n! ~2^d}}\g_{\mu_1 ...\mu_d}^{\a\b} ~(\l^\r \g^{\mu_1 ...\mu_d}_{\r\de}
\l^\de)
\end{equation}
where
$\l\g^{\mu_1 ... \mu_d} \l$ defines an $d$-dimensional complex plane $\mathbb{C}^{d} \subset \mathbb{R}^{2d} \otimes \mathbb{C}$.
This
$d$-dimensional complex plane is preserved  by a $U(d)$
subgroup of $SO(2d)$
rotations. Also, multiplying $\l$ by a non-zero complex number does not change this plane. So, if we consider the space of $\l$'s obeying  up to rescalings, the space of {\it projective pure spinors}, $\mathbb{P}{\rm PS}_{2d}$ in $D=2d$ Euclidean dimensions, then:
\begin{equation}
\mathbb{P}{\rm PS}_{2d} = SO(2d)/U(d)
\end{equation}
The real dimension of this space is $d (d-1)$. The space ${\rm PS}_{2d}$ of pure spinors is a cone over $\mathbb{P}{\rm PS}_{2d}$. 
The space $X_{2d}$, which is ${\rm PS}_{2d}$ with the point $\l = 0$ deleted, can be thought of the moduli space of Calabi-Yau complex structures on $\mathbb{R}^{2d}$, i.e. the space of pairs 
$$
( {\rm identification} \  \mathbb{C}^{d} \approx \mathbb{R}^{2d} , {\Omega} \in {\Lambda}^{d} \mathbb{C}^{d} )
$$
This is an important space in the context of B type topological strings.

\subsection{Pure Spinor Parameterization}\label{PSP}

In order to solve the 10-dimensional  pure spinor constraints it is useful to write them in terms of the $U(5)$ variables.

A vector in 10-dimensions, $\tilde V^\mu$,  can be written as a direct sum of two 5-dimensional  vectors
\begin{align}
V^a &:= \frac{1}{\sqrt{2}} (\tilde V^a +i \tilde V^{a+5}), \qquad  a=1,2,...,5\\
V_a &:= \frac{1}{\sqrt{2}} (\tilde V^a -i \tilde V^{a+5}),
\end{align}
i.e. we have broken the 10-dimensional vector representation of $SO(10)$ as a sum of two vectorial representations of $U(5)$, $10 = 5 \oplus \bar 5$. In the 10-dimensional Gamma matrices  we have
\begin{align}
b^a &:= \frac{1}{\sqrt{2}} (\G^a +i \G^{a+5}), \qquad  a=1,2,...,5\\
b_a &:= \frac{1}{\sqrt{2}} (\G^a -i \G^{a+5}),
\end{align}
where the Gamma-algebra becomes $\{b_a,b^c\}=\de_a^c$.  Now, the $(b_a, b^c)$ matrices satisfy a ladder algebra and we can construct a finite representation.

We define the fundamental state such that $b_a|0\rangle =0, a=1,...,5$, so all states are created applying the $b^\a$ matrix on $|0\rangle$. Since that the pure spinor is a chiral spinor and the chiral operator just counts the number of $b^a$ matrices which acts on $|0\rangle$, then the most general positive chiral spinor is written as
\begin{equation}
|\l^\a\rangle = \l^+ |0\rangle + \frac{1}{2}\l_{ab}b^a b^b |0\rangle + \frac{1}{24} \l^a \e_{abcde} b^b b^c b^d b^e |0\rangle,
\end{equation}
where positive chirality means the number of $b^a$ is even and $\l_{ab} = -\l_{ba}$. Clearly, we have broken the $\l^\a$ spinor as $\l^\a = (\l^+, \l_{ab}, \l^a)$, where the number of degrees of freedom of $\l^+$ is one, of $\l_{ab}$ is 10 and the $\l^\a$ is 5, namely $16 \rightarrow (1,\bar 10,5)$.

Finally, using the $U(5)$ representation the pure spinor constraints becomes
\begin{align}
\l^+\l^a + \frac{1}{8}\e^{abcde}\l_{bc} \l_{de} &=0,   \qquad  a=1,...,5, \label{eq1}\\
\l^b \l_{ba} &=0 \label{eq2}. 
\end{align}
Choosing the chart where $\l^+\neq 0 $ and using the parameterization $\l^+=\g, \, \l_{ab}=\g\, u_{ab}$, the solution of the equations in \eqref{eq1} is straightforward
\begin{equation}
\l^a =-  \frac{\g}{8}\e^{abcde}u_{bc} u_{de},
\end{equation}
and the equations in \eqref{eq2} becomes trivial.

As a final remark, because the pure spinor has ghost number 1, then obviously  $\g$ has ghost number 1 and $u_{ab}$ has ghost number 0. Therefore, we can write an holomorphic top form over the pure spinor space with ghost number $8$ as
\begin{equation}
[d\l^\a] = \g^7 \,d\g\wedge du_{12}\wedge du_{13}\wedge\cdots \wedge du_{45},
\end{equation}
which matches with the one written in  \eqref{top-form}.

%%%%%%%%%%%%%%%%%%%%%%%%%%%%%%%%%%%%%%%%%%%%%%%%%%%%%%%%%%%%%%

\end{document}